\documentclass[onecolumn, 12pt]{IEEEtran}

\ifCLASSINFOpdf
   \usepackage[pdftex]{graphicx}
   \graphicspath{{../pdf/}{../jpeg/}}
   \DeclareGraphicsExtensions{.pdf,.jpeg,.png}
\else
   \usepackage[dvips]{graphicx}
   \graphicspath{{../eps/}}
   \DeclareGraphicsExtensions{.eps}
\fi
\usepackage{diagbox}
\usepackage{color}
\usepackage{amsmath}
\usepackage{bm}
\usepackage{amssymb}
\usepackage{multirow}   
\usepackage{makecell}
\usepackage{setspace}
\begin{document}
%
\title{AnciNet: An Efficient Deep Learning Approach for Feedback Compression of Estimated CSI in Massive MIMO Systems}
%
%
%

\author{Yuyao~Sun,~\IEEEmembership{Student Member,~IEEE,}
        Wei~Xu,~\IEEEmembership{Senior Member,~IEEE,}
        Lisheng~Fan,~\IEEEmembership{Member,~IEEE,}
        Geoffrey~Ye~Li,~\IEEEmembership{Fellow,~IEEE,}
        and~George~K.~Karagiannidis,~\IEEEmembership{Fellow,~IEEE}
\thanks{Y. Sun is with the National Mobile Communications Research Laboratory, Southeast University, Nanjing 210096, China (e-mail: yy.sun@seu.edu.cn).}
\thanks{W. Xu is with the National Mobile Communications Research Laboratory, Southeast University, Nanjing 210096, China, and also with Purple Mountain Laboratories, Nanjing 211111, China (e-mail: wxu@seu.edu.cn).}
\thanks{L. Fan is with the School of Computer Science, Guangzhou University, Guangzhou 510006, China (e-mail: lsfan@gzhu.edu.cn).}
\thanks{G. Y. Li is with the School of Electrical and Computer Engineering, Georgia Institute of Technology, Atlanta, GA, USA (e-mail: liye@ece.gatech.edu).}
\thanks{G. K. Karagiannidis is with the Electrical and Computer Engineering Department, Aristotle University of Thessaloniki, 54 124 Thessaloniki, Greece (e-mail: geokarag@auth.gr).}}

\maketitle

\begin{abstract}
Accurate channel state information (CSI) feedback plays a vital role in improving the performance gain of massive multiple-input multiple-output (m-MIMO) systems, where the dilemma is excessive CSI overhead versus limited feedback bandwith. By considering the noisy CSI due to imperfect channel estimation, we propose a novel deep neural network architecture, namely AnciNet, to conduct the CSI feedback with limited bandwidth. AnciNet extracts noise-free features from the noisy CSI samples to achieve effective CSI compression for the feedback. Experimental results verify that the proposed AnciNet approach outperforms the existing techniques under various conditions.
\end{abstract}


%
\IEEEpeerreviewmaketitle

\vspace{2cm}
\section{Introduction}
%
%
%
%
\IEEEPARstart{M}{assive} multiple-input multiple-output (m-MIMO) has been a fundamental component of the 5G wireless network. However, to fully exploit this technology, accurate channel state information (CSI) must be acquired at the transmitter. In frequency division duplex (FDD) systems, downlink CSI is generally estimated at user equipment (UE) and then fed back to the base station (BS). However, the channel matrix in m-MIMO systems is huge due to the large number of antennas, which makes CSI estimation and feedback very challenging, especially through a bandwidth-limited feedback channel.

The urgent demand to reduce the CSI feedback has motivated various techniques, such as codebook-based \cite{ref1} and compressive sensing (CS)-based \cite{ref2}. The codebook-based method quantizes the CSI into an index of a codeword in a predetermined codebook while the overhead of the method scales linearly with the number of antennas. The CS-based method operates iteratively to transform the CSI matrix into certain base.

Recently, the successful application of deep learning (DL)-based methods in channel estimation and signal detection \cite{ref3} has inspired a series of works \cite{ref4}-\cite{ref6}. This also attracted increasing attention for m-MIMO CSI feedback. An autoencoder-based neural network (NN), named CsiNet, has been proposed in \cite{ref7} to learn the process of CSI feedback. Specifically, the encoder compresses CSI at the UE and then the decoder at the BS is utilized to restore the channel matrix from the compressed representation. Then, an extended version, namely CsiNet+, has been developed in \cite{ref8} to further improve the performance of CsiNet by enlarging the convolutional kernel size. In \cite{ref9}, a joint convolution residual network is used to optimize CSI quantization for feedback. In \cite{ref10} and \cite{ref11}, the CSI compression and reconstruction are improved by exploiting the temporal correlation of CSI in adjacent time slots through an NN module, named long short-term memory (LSTM). In \cite{ref12}, the uplink-downlink channel reciprocity is utilized to assist downlink CSI estimation. Furthermore, in \cite{ref13}, the CRBlock, consisting of multi-resolution paths, is introduced to improve the robustness of NN-based CSI feedback over different compression ratios (CRs). Nevertheless, the above NN-based methods have been devised by presuming perfect CSI, which however can rarely be achieved in practice. Therefore, it is expected to have a CSI feedback compression method that takes the noisy CSI into consideration.

In this article, we develop a deep NN, namely anti-noise CSI compression network (AnciNet), for noisy CSI feedback compression. Specifically, we propose the novel NN architecture, AnciNet, to deal with the problem of noisy CSI feedback compression, which comprises of two new modules, i.e., a pre-denoising module and an Anci-block enhanced feedback module. The pre-denoising module at the front-end of the NN preliminarily suppresses noise before encoding while the Anci-block enhanced feedback module addresses the residual noise during the compression and decompression process. With both functional modules, the proposed NN can restore the CSI from a noisy input with high accuracy at the other end. In particular, the structure of our proposed Anci-block can effectively eliminate the noise of the m-MIMO CSI by placing two small convolutional layers parallelly, which are used to post-process data from the preceding convolutional layer with a larger kernel size.

\section{System Model}
We consider the downlink of an FDD m-MIMO system with $N_t$ antennas at the BS and a single-antenna at the UE. Orthogonal frequency division multiplexing (OFDM) with ${N_c}$ subcarriers is employed. The received signal at the $n$th subcarrier can be expressed as
\begin{equation}
{y_n} = {w_n}{x_n} + {z_n}, \label{1}
\end{equation}
where ${x_n} \in \mathbb{C}$ and ${z_n} \in \mathbb{C}$ are the transmit symbol and additive noise at the $n$th subcarrier, respectively, and $w_n  \buildrel \Delta \over = {\bf{h}}_n^H {\bf{v}}_n$ is the equivalent channel coefficient, where ${{\bf{h}}_n} \in {\mathbb{C}^{{N_t} \times 1}}$ and ${{\bf{v}}_n} \in {\mathbb{C}^{{N_t} \times 1}}$ denote the channel vector and precoding vector corresponding the $n$th subcarrier, respectively.

In the spatial-frequency domain, the downlink channel matrix is denoted by ${\bf{H}} = {[{{\bf{h}}_1} \cdots {{\bf{h}}_{{N_c}}}]^H} \in {\mathbb{C}^{{N_c} \times {N_t}}}$. Without compression, the number of total feedback parameters, i.e., the size of ${\bf{H}}$, is ${N_c}{N_t}$, which is extremely large in m-MIMO systems. Hence, we have to compress CSI for feedback over limited bandwidth resource. Since channel matrix, ${\bf{H}}$, is sparse in the angular-delay domain \cite{ref14}, we can compress the channel by transforming ${\bf{H}}$ from the spatial-frequency domain to the angular-delay domain. Using a 2D discrete Fourier transform (DFT), the channel matrix in the angular-delay domain can be expressed as
\begin{equation}
{{\bf{H}}_d} = {{\bf{F}}_c}{\bf{H}}{\bf{F}}_t^H, \label{2}
\end{equation}
where ${{\bf{F}}_c}$ and ${{\bf{F}}_t}$ represent two DFT matrices with dimensions ${N_c} \times {N_c}$ and ${N_t} \times {N_t}$, respectively. Based on the fact that multipaths arrive at limited delay intervals \cite{ref9}, ${{\bf{H}}_d}$ contains only values in a small delay duration. Without loss of generality, we focus on ${{\bf{H}}_s} \in {\mathbb{C}^{{N_{cc}} \times {N_t}}}$, the first ${N_{cc}}$ rows of ${{\bf{H}}_d}$ in the angular-delay domain.

Due to the imperfection of  channel estimation, noisy CSI is available. By following a typical linear model of the CSI error \cite{ref15}, the noisy version of the shortened channel matrix can be expressed as
\begin{equation}
\widehat{\bf{H}}_s = {{\bf{H}}_s} + {\bf{E}}, \label{3}
\end{equation}
where ${\bf{E}} \in {\mathbb{C}^{{N_{cc}} \times {N_t}}}$ represents the additive noise.

Although the number of the feedback parameter decreases from ${N_c}{N_t}$ to ${N_{cc}}{N_t}$, further compression is still necessary since ${N_{cc}}{N_t}$ is still too large for m-MIMO. To this end, we design a DL-based approach consisting of a denoising module and an autoencoder-based feedback module to achieve effective feedback and recovery of CSI from a noisy version of the CSI estimate. More specifically, as shown in Fig. 1(a), the denoising module provides a preliminary noise reduction operation to obtain
\begin{equation}
\widetilde{\bf{H}}_s = {f_{DN}}({\widehat{\bf{H}}_s}), \label{4}
\end{equation}
a cleaner input for the latter module. The feedback module includes an Encoder, ${\bf{s}} \buildrel \Delta \over = f_{EN} (\widetilde{\bf{H}}_s )$, and a Decoder, ${\bf{H}}_{de} = {f_{DE}}({\bf{s}})$, which is responsible for the CSI compression and decompression to minimize the impact of residual noise from the denoising module.

\begin{figure}[t]
\centering
\includegraphics[width=0.75\textwidth]{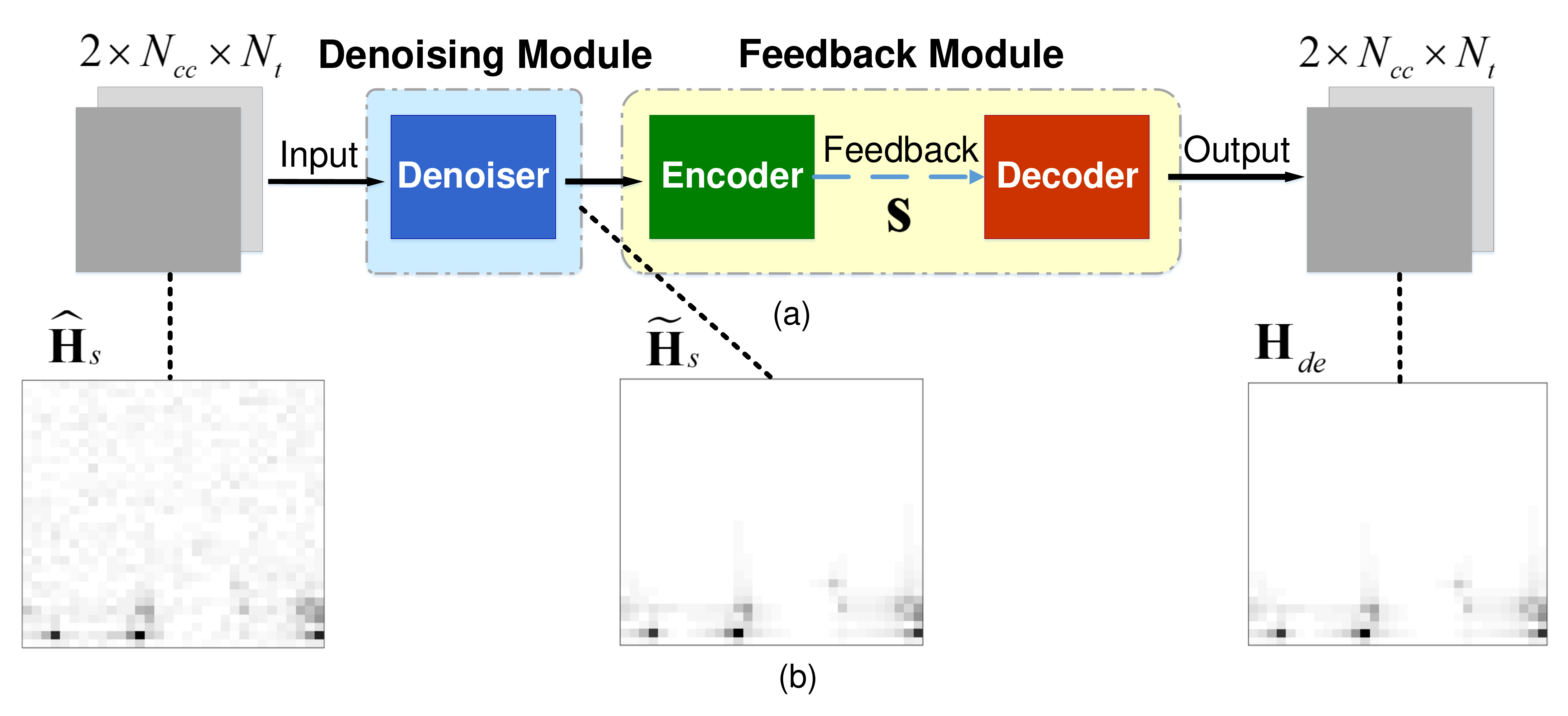}
\caption{(a) Architecture of AnciNet. (b) Description samples of $\widehat{\bf{H}}_s$, $\widetilde{\bf{H}}_s$, ${\bf{H}}_{de}$.}
\end{figure}

For better illustration, we depict in Fig. 1(b) an instance of CSI samples of $\widehat{\bf{H}}_s$, $\widetilde{\bf{H}}_s$ and ${\bf{H}}_{de}$ generated with the COST 2100 indoor channel model \cite{ref16}, where the strength of a pixel represents the magnitude of a channel gain in the angular-delay domain. The Encoder imposes a cutdown on feedback overhead via compressing the channel matrix, ${\widetilde{\bf{H}}_s}$, into a low-dimensional ($M$-dimensional) codeword, ${\bf{s}}$. Then, the number of the total feedback parameters reduces to $M$. We use two real-valued matrices to denote the real and imaginary parts of ${\widehat{\bf{H}}_s}$ for real NNs. The corresponding CR will be $\gamma  = M/2{N_{cc}}{N_t}$. As an inverse of the Encoder, the Decoder reconstructs the channel matrix from the codeword, ${\bf{s}}$.

\section{AnciNet and Training}
In this section, we present the architecture of the proposed AnciNet, including the denoising and the feedback modules as well as its training method.

\subsection{Denoising Module with Anci-block}
Convolutional neural network (CNN)-based DL techniques have been widely applied in the field of image denoising \cite{ref17}, \cite{ref18}. By considering the CSI matrix as a 2-D image, these techniques can be utilized. However, there are some differences between the denoising of CSI images and that of common pictures. The most important difference is that the CSI matrix is approximately sparse in the angular-delay domain. Therefore, it is critical to take this advantage when designing the denoising module.

Inspired by \cite{ref18}, we devise the Anci-block in the AnciNet, to extract noise-free features of the noisy CSI input. Fig. 2 illustrates the details of AnciNet architecture, where the three-dimensional size values, e.g., $(2  \times {N_{cc}}  \times {N_t} )$, on the top of each block denote the depth, length, and width of the corresponding input tensor, respectively. The four-dimensional values, e.g., $(64 \times 2  \times 2  \times 7 )$, denote the number, depth, length, and width of the convolution kernel, respectively. As depicted in the dashed yellow box of Fig. 2, the Anci-block consists of three composite units. Each unit is a proper combination of a convolutional layer, a batch normalization layer, and an activation layer using the Leaky Rectified Linear Unit (Leaky ReLU) function
\begin{equation}
{\rm{Leaky ReLU}}(x) = \left\{ \begin{array}{l}
 x, \quad \ \, x \ge 0, \\
 0.3x, \;  x < 0 .\\
 \end{array} \right.      \label{5}
\end{equation}
The convolutional layer in the first composite unit of an Anci-block uses kernels of size, $16 \times 7 \times 7$, to generate 16 feature maps. Considering that the CSI images are not as complex as traditional common colourful pictures, 16 feature maps are sufficient to catch the features. Zero padding is applied to keep the length and width of the output tensor same as the input tensor. Then the output of the first composite unit passes through the two parallel composite units with kernel sizes, $16 \times 5 \times 5$ and $16 \times 3 \times 3$. At the end of each Anci-block, features captured by these two units are additively combined.

A key structure of the Anci-block is to place the convolutional layers with small kernel sizes (the last two dimensions, e.g., $3 \times 3$, $5 \times 5$) behind the convolutional layer with a larger kernel size (e.g., $7 \times 7$). Thanks to the larger size of receptive field, the larger kernel has shown excellence in estimating noise-free features, thereby eliminating the noise effectively. While there is a side effect that the larger kernel also smoothens the image details. This effect is compensated in the Anci-block by adding these small kernels that are capable of extracting fine features but simultaneously capturing noise. Therefore, by placing the convolutional layers with proper small kernel sizes behind the convolutional layer with a larger kernel size, noise-free details can be effectively extracted from the noisy input.

\begin{figure}[t]
\centering
\includegraphics[width=0.75\textwidth]{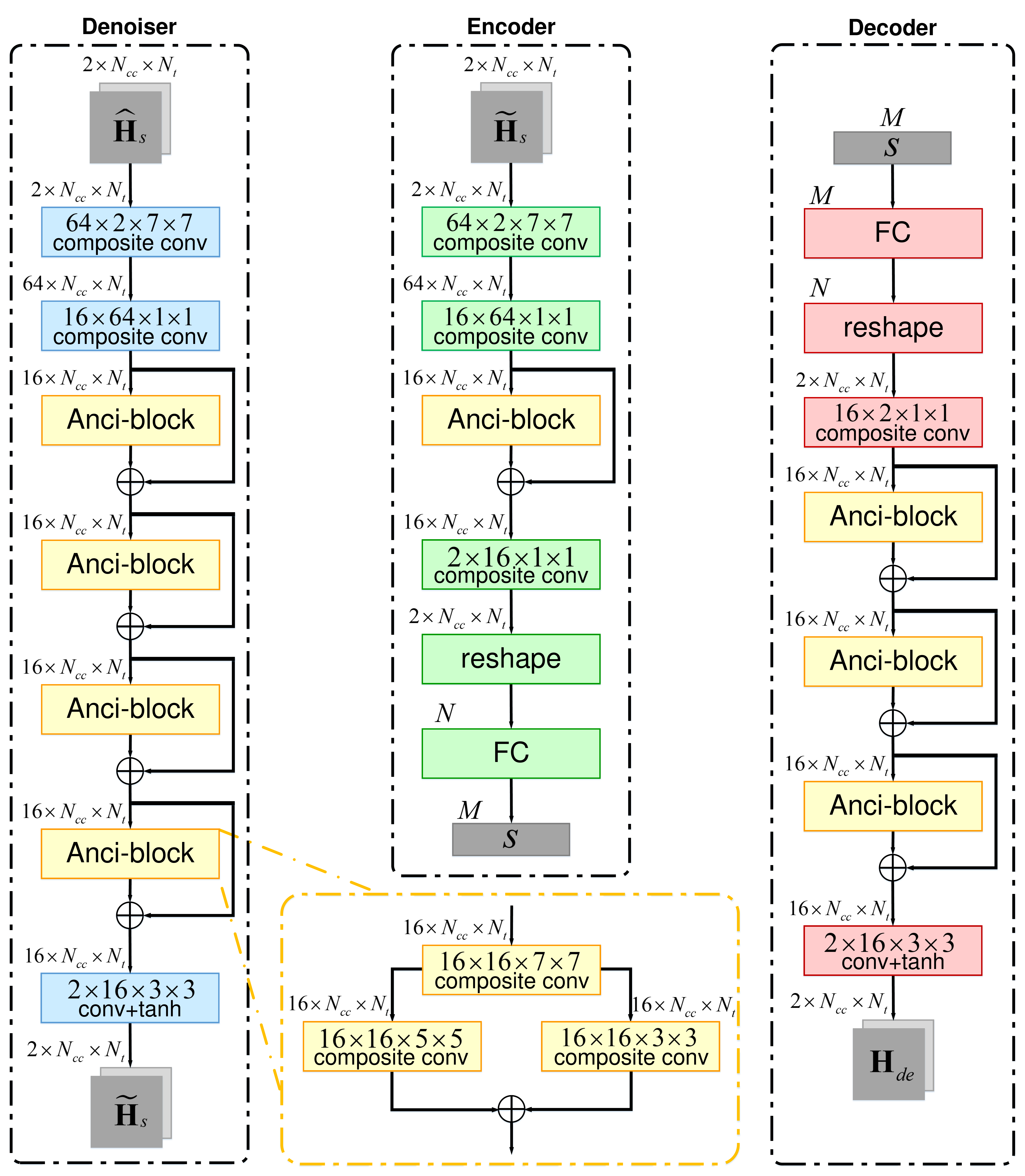}
\caption{Architecture of the Denoiser, Encoder and Decoder.}
\end{figure}

The above design of Anci-block makes use of the specific characteristics of the CSI image. Its design philosophy is elaborated with the help of Fig. 3. In Fig. 3, a typical CSI sample from the COST 2100 indoor channel model \cite{ref16} is used. It should be noted that the CSI image in the angular-delay domain usually composed of several clusters and each includes a center and ``insignificant" details, as we can see from the three clusters in Fig. 3. The center of a cluster in practice corresponds to a resolvable path in the m-MIMO channel. Thus it has relatively much higher magnitude and in general occupies a compact and small area, which makes it reasonable to view these centers as dominate details of the CSI image. The rest of the pixels, excluding the ``center", in a cluster of the CSI image in Fig. 3 correspond to low-power propagation paths near the strong ``center" path. These pixels can be regarded as ``insignificant" details of the CSI image for two reasons. Firstly, they are less significant and even more noisy compared to the centers. Secondly, the CSI image except for the clusters is nearly zero valued and should not be ignored, even though less significant.

\begin{figure}[t]
\centering
\includegraphics[width=0.75\textwidth]{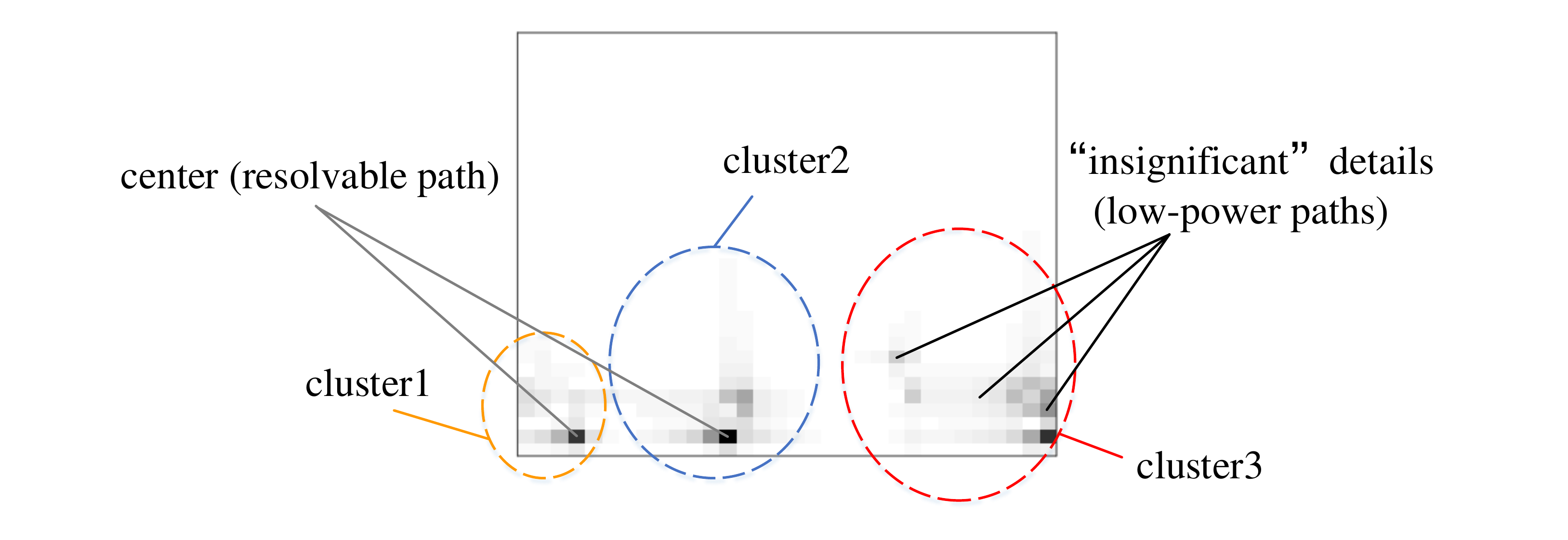}
\caption{Illustration of the cluster, center and ``insignificant" details.}
\end{figure}

Upon considering the above unique features of the CSI image, the first composite unit in an Anci-block in our proposed design is responsible for reducing the general noise and the next two parallel composite units extract deeper features of the details. Specifically, the smaller $3 \times 3$ kernel extracts the subtle details of the high-magnitude compacted ``center" and the $5 \times 5$ kernel extracts the features of the low-magnitude wide-range ``insignificant" details. As illustrated in the left column of Fig. 2, by stacking four Anci-blocks, the Denoiser exhibits an appealing ability of eliminating the noise while reserving subtle features. The first block of the Denoiser is a composite unit, which utilizes $2 \times 7 \times 7$ convolution to generate 64 feature maps. Here we use a number of large filters to exploit multiple and profound features before further processing. Then a composite unit with the filter size of $64 \times 1 \times 1$ is employed to make the input size same as the output size of the next Anci-block. Residual learning \cite{ref19} is applied to deal with the problem of performance degradation caused by vanishing gradient. After four cascaded Anci-blocks, a preliminarily denoised CSI image, ${\widetilde{\bf{H}}_s}$, is reconstructed through the $16 \times 3 \times 3$ convolutional layer with ``tanh" as the activation function.

\subsection{Anci-block Enhanced Feedback Module}
From Fig. 1(a), the feedback module consists of an Encoder and a Decoder. The feedback module is designed according to the principle that residual noise ought to be minimized during the connected compression and decompression processes, thus the Anci-block is also used in the Encoder and Decoder architecture.

The center column of Fig. 2 specifies the architecture of Encoder. The first two blocks of Encoder play a similar role to those of Denoiser. An Anci-block is then inserted to generate clean feature maps, after which a $16 \times 1 \times 1$ composite unit compresses the feature volume to $2 \times {N_{cc}} \times {N_t}$. At the end of the Encoder, a fully-connected layer is used to further compress the reshaped vector into codeword, ${\bf{s}}$.

The structure of Decoder is demonstrated at the right column in Fig. 2. Once the codeword, ${\bf{s}}$, is received, a fully connected layer is first adopted to demap the compressed features into a higher $N$-dimensional vector, which provides not only the restoration of its original size but also a preliminary estimate of the ground true values of ${{\bf{H}}_s}$. In order to enhance the performance of CSI reconstruction, we use a $2 \times 1 \times 1$ composite unit to increase the number of feature maps after resizing the vector. Then three concatenated Anci-blocks with residual learning exploits the noise-free features. Finally, a $16 \times 3 \times 3$ convolutional layer with ``tanh" activation function is introduced to obtain a fully recovered channel matrix, ${\bf{H}}_{de}$.

\subsection{Two-Stage Training}
Despite the fact that the denoising module and the feedback module are placed in a sequential manner, they expect to play different roles. The feedback module is in charge of CSI compression and decompression whereas the denoising module focuses more on noise eliminating. Therefore, these two modules in our proposed AnciNet ought to be trained separately instead of popular end-to-end training adopted by most of the existing CsiNet-like NN, e.g., CsiNet-LSTM \cite{ref10}.

In order to train AnciNet, we devise a two-stage training approach to maximize the individual performance metric of each module, which will in turn achieve the optimal results of the entire cascaded network. In the first stage, we train the denoising module independently. $\widehat{\bf{H}}_s [i]$ denotes the $i$th sample of $\widehat{\bf{H}}_s$ and the network is trained to minimize the mean-squared error (MSE) between the output of the denoiser, ${f_{DN} (\widehat{\bf{H}}_s[i] )}$, and ${\bf{H}}_s[i]$. Thus the loss function will be
\begin{equation}
L_1  = \frac{1}{T_1}\sum\limits_{i = 1}^{T_1} {\left\| {f_{DN} (\widehat{\bf{H}}_s[i] ) - {\bf{{H}}}_s[i] } \right\|_2^2 },   \label{6}
\end{equation}
where $T_1$ represents the total number of training samples and $\left\|  \cdot  \right\|_2$ takes the Euclidean norm.

After the first stage, the training samples for the second stage are obtained from the well-trained output of the denoiser, $\widetilde{\bf{H}}_s$. The reconstructed CSI is ${\bf{H}}_{de}[i]  = f_{DE} (f_{EN} (\widetilde{\bf{H}}_s[i] ) )$. We also use the MSE loss function to train the network, which can be expressed as
\begin{equation}
L_2=\frac{1}{T_2}\sum\limits_{i = 1}^{T_2 }{\left\| {f_{DE} (f_{EN} (\widetilde{\bf{H}}_s[i]  ) ) - {\bf{H}}_s[i] } \right\|_2^2 },
\end{equation}
where $T_2$ represents the total number of training samples.

\section{Experimental Results}
This section conducts some experimental comparisons of the proposed AnciNet and existing methods for CSI compression feedback using deep NN.

\subsection{Experiment Parameters}
We generate a total of 150,000 samples of channel realizations with the COST 2100 indoor channel model \cite{ref16} at 5.3 GHz. The BS is equipped with a uniform linear array (ULA) with $N_t=32$ antennas and we set $N_c=1024$ subcarriers. After transforming into the angular-delay domain, the first $N_{cc}=32$ rows of ${\bf{H}}_d$ in the delay domain are kept. Then additive noise, ${\bf{E}}$, is added to simulate the error of imperfect channel estimation. Note that the proposed method does not have any restriction on the distribution of ${\bf{E}}$, which may come from a number of sources, e.g., channel estimation error and hardware impairments. $\widehat{\bf{H}}_s$ is normalized to $( - 1,1)$ for better data processing optimization. The average power of ${\bf{H}}_s$ to that of ${\bf{E}}$ ratio is denoted as the channel-to-noise ratio (CNR). For a fixed CNR, the training, validation and test sets contain 100,000, 30,000, 20,000 samples of $\widehat{\bf{H}}_s$ and their corresponding labels, ${\bf{H}}_s$.

The simulation is carried out in Keras on an NVIDIA GTX1080Ti GPU. For both of the two stages, we adopt Adam optimizer to update the training parameters. The epochs are set to be 1,000 and the batch size is 1,000.

\subsection{Performance Evaluation}
We use normalized MSE (NMSE) to evaluate the performance of CSI estimation of the proposed AnciNet, which is defined as
${\rm{NMSE}} = E\left\{ {{\left\| {{\bf{H}}_{de}  - {\bf{H}}_s } \right\|_2^2 }}/{{\left\| {{\bf{H}}_s } \right\|_2^2 }} \right\}$.
\begin{figure}[t]
\centering
\includegraphics[width=0.75\textwidth]{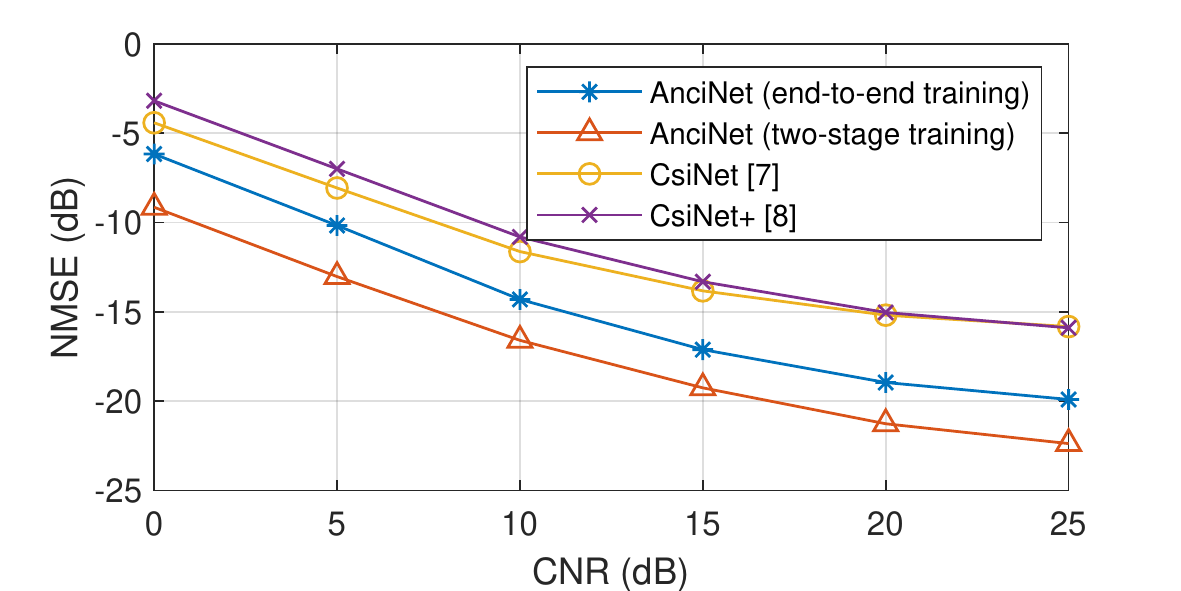}
\caption{NMSE performance at different CNRs.}
\end{figure}
\begin{table}
\begin{center}
\caption{Number of Parameters}
\begin{tabular}{|c|c|c|c|c|}
\hline
\diagbox{Method}{$\gamma$}& 1/4 & 1/16 & 1/32 & 1/64\\ 
\hline
 CsiNet [7] & 2,103,904 & 530,656 & 268,448 & 137,344\\
\hline
 AnciNet & 2,289,334 & 716,086 & 453,878  & 322,774\\
\hline
\end{tabular}
\end{center}
\end{table}

Regarding the training complexity of AnciNet, the convolutional layers and the fully-connected layers occupy a majority of the total parameters. Specifically, the parameters of a four-dimensional convolutional layer with kernel size $f \times k \times m \times m$ is  $n_c = f \times k \times m \times m + f$ and the parameters of a fully-connected layer is $n_f = f_{\rm{in}} \times f_{\rm{out}} + f_{\rm{out}}$, where $f_{\rm{in}}$ is the number of input neurons of the fully-connect layer and $f_{\rm{out}}$ is the number of output neurons, respectively. The parameter size of AnciNet is listed in Table I, which is comparable with that of existing NNs, e.g., CsiNet \cite{ref7}.

Fig. 4 compares the NMSE performance achieved by CsiNet \cite{ref7}, CsiNet+ \cite{ref8}, AnciNet with the two-stage training, and AnciNet with the end-to-end training at different CNRs with $\gamma  = 1/4$. All of the three NNs are trained with dataset at ${\rm{CNR}} = 10$ dB. For comparison, we replace the activation function ``sigmoid" with ``tanh" in CsiNet and CsiNet+. Both AnciNets outperform CsiNet and CsiNet+ thanks to the proposed Anci-block. However, there is a certain disparity between the two AnciNets trained with different approaches. To be specific, the two-stage trained AnciNet delivers better performance than the end-to-end trained AnciNet, especially when CNR is lower than 5 dB. This evidences our motivation of inventing the two-stage training procedure for the proposed AnciNet that can boost network training performance.
\begin{figure}[t]
\centering
\includegraphics[width=0.75\textwidth]{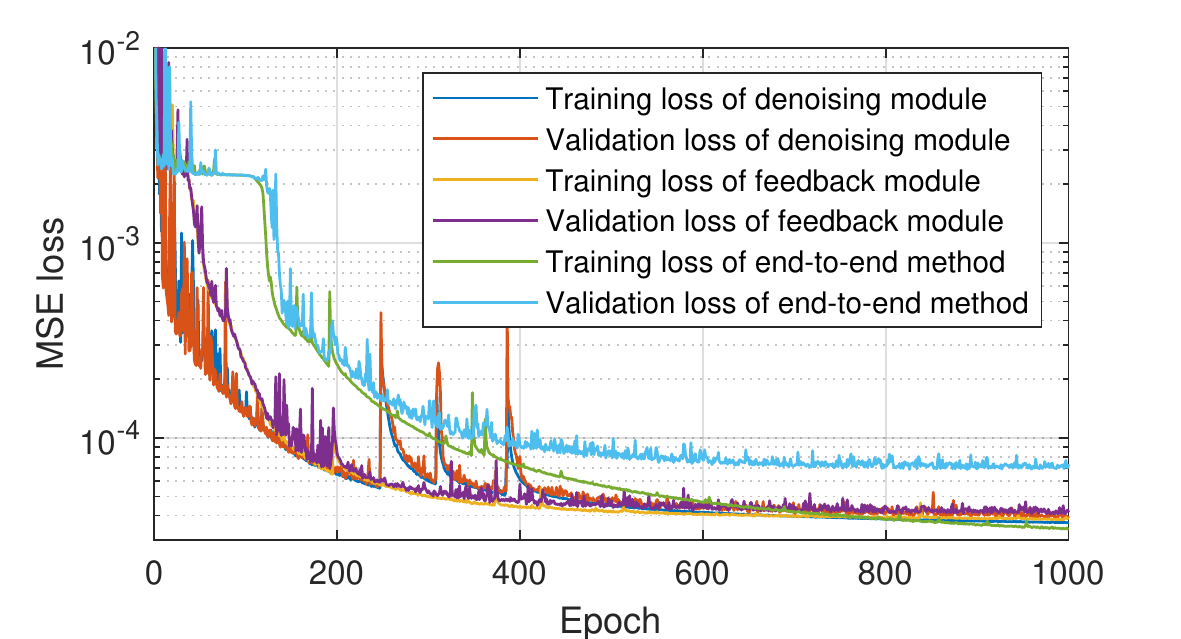}
\caption{Loss comparison between the two training methods.}
\end{figure}
\begin{table}[t]
\caption{NMSE Comparison at Different CRs and CNRs}
\begin{center}
\begin{tabular}{p{0.08\textwidth}<{\centering}|p{0.12\textwidth}<{\centering}|p{0.13\textwidth}<{\centering}|p{0.13\textwidth}<{\centering}}
\hline
\multirow{2}{*}{$\gamma$} &
\multirow{2}{*}{CNR(dB)} &
\multicolumn{2}{c}{NMSE(dB)} \\
\cline{3-4}   
& &CsiNet [7] &AnciNet \\
\hline
\hline
\multirow{6}{*}{$\frac{1}{16}$} & 0 &-2.76 &-8.70\\
&5 &-6.24 &-11.64\\
&10 &-8.71 &-13.31\\
&15 &-9.56 &-14.42\\
&20 &-10.03 &-14.84\\
&25 &-10.19 &-14.99\\
\hline
\multirow{6}{*}{$\frac{1}{32}$} & 0 &-1.94 &-7.47\\
&5 &-4.94 &-9.26\\
&10 &-7.23 &-9.88\\
&15 &-7.83 &-10.42\\
&20 &-8.09 &-10.55\\
&25 &-8.17 &-10.60\\
\hline
\multirow{6}{*}{$\frac{1}{64}$} & 0 &-0.75 &-6.49\\
&5 &-3.45 &-7.65\\
&10 &-4.80 &-7.92\\
&15 &-5.28 &-8.30\\
&20 &-5.45 &-8.37\\
&25 &-5.52 &-8.40\\
\hline
\end{tabular}
\end{center}
\end{table}

Fig. 5 elaborates the reason of the superiority of the two-stage training approach. The descending trend of the training and validation loss indicates that although the AnciNet trained with an end-to-end method can reach a training loss no worse than the two-stage trained AnciNet after 600 epochs, it suffers overfitting after 300 epochs and leads to worse individual performance of each module. On the contrary, the two modules trained separately rarely suffers the problem of overfitting, thus achieving both better individual and entire performance.

For comparison, we also summarize the NMSE performance of CsiNet and the two-stage trained AnciNet at different CRs and CNRs in Table II. Both approaches are trained with the same dataset at ${\rm{CNR}} = 10$ dB. From Table II, the proposed AnciNet outperforms CsiNet at different CNRs and CRs. In particular, the performance of CsiNet degrades noticeably with CR lower than $1/32$ while AnciNet is still able to reconstruct the channel matrix in a much more accurate manner. On the other hand, when the CNR is lower than 5 dB, the proposed AnciNet shows its robustness in CSI recovery at ${\rm{CNR}} = 0$ dB while CsiNet can hardly satisfy there. AnciNet performs well at low-to-moderate speeds. For high-speed environments, further enhancements by involving more sophisticated techniques, like LSTM \cite{ref11}, are required.

\section{Conclusion}
In this article, we have proposed a novel NN called AnciNet for noisy CSI compression and feedback in m-MIMO systems. We have also introduced a two-stage training method to enhance training progress. Simulation results have shown that AnciNet performs well at various conditions and can be used for practical scenarios.

\ifCLASSOPTIONcaptionsoff
  \newpage
\fi

%

%






\end{document}